# Suppressing parametric resonance of a Hyperloop vehicle using a parametric force

Jithu Paul[a,*], Karel N. van Dalen[a], Andrei B. Fărăgău[a], Rens J. van Leijden[a], Mouad Ouggaâli[a], Andrei V. Metrikine[a]

[a]*Department of Engineering Structures, Faculty of CEG, TU Delft, NL*

In this paper, we study the stability of a simple model of a Hyperloop vehicle resulting from the interaction between electromagnetic and aeroelastic forces for both constant and periodically varying coefficients (i.e., parametric excitation). For the constant coefficients, through linear stability analysis, we analytically identify three distinct regions for the physically significant equilibrium point. Further inspection reveals that the system exhibits limit-cycle vibrations in one of these regions. Using the harmonic balance method, we determine the properties of the limit cycle, thereby unravelling the frequency and amplitude that characterize the periodic oscillations of the system's variables. For the varying coefficients case, the stability is studied using Floquet analysis and Hill's determinant method. The part of the stability boundary related to parametric resonance has an elliptical shape, while the remaining part remains unchanged. One of the major findings is that a linear parametric force can suppress or amplify the parametric resonance induced by another parametric force depending on the amplitude of the former. In the context of the Hyperloop system, this means that parametric resonance caused by base excitation—in other words by the linearized parametric electromagnetic force—can be suppressed by modulating the coefficient of the aeroelastic force in the same frequency. The effectiveness is also highly dependent on the phase difference between the modulation and the base excitation. The origin of the suppression is attributed to the stabilizing character of the parametric aeroelastic force as revealed through energy analysis. We provide analytical expressions for the stability boundaries and for the stability's dependence on the phase shift of the modulation. Finally, we emphasize that suppressing parametric resonance through an added, linear state-dependent force with coefficient having the same period as the original force can be achieved in other physical systems too.

Keywords: Hyperloop, electro-magnetic suspension, Floquet theory, Hill's determinant method, harmonic balance, limit cycle, suppression of parametric resonance, aeroelastic force, supercritical Hopf bifurcation, interaction of state-dependent forces.

## 1 Introduction

The Hyperloop is expected to revolutionize transportation, blending the advantages of aircraft and next-generation rail. This unique fusion yields a richer engineering landscape, presenting an open field for research. While the aeroelastic stability of aircraft and the wave-induced instability (related to the flexible guideway) of conventional rail have been studied rather extensively, the combination with magnetic levitation employed in modern rail systems remains an open area for exploration. In the context of a Hyperloop vehicle traveling within a depressurized tube, levitated electromagnetically from a flexible beam, the potential for integrating the aforementioned mechanisms (aeroelastic, electromagnetic and wave-induced) arises. However, whether these stability mechanisms complement or conflict with each other remains to be seen. Noteworthy literature pertaining to each individual mechanism is cited below.

It is widely recognized that when a vehicle move along a flexible guideway, oscillations can become unstable if its speed exceeds a specific critical threshold [1]. Metrikine [2] demonstrated that instability arises due the radiation of anomalous Doppler waves, which feedback energy into the vehicle's vibration, surpassing that of normal Doppler waves. Identifying the critical velocity beyond which this may happen is imperative in the design phase [3]. Paddison et. al. [4] studied the control implications of magnetically-suspended vehicles having relatively soft chassis structures.

The primary aeroelastic effects that could impact a Hyperloop vehicle include galloping [5], fluttering, and vortex-induced vibrations [6,7], although they overlap to some extent. In this paper we mainly focus on galloping which can be characterized as being a low-frequency instability phenomenon of aerodynamic nature, and it usually occurs on slender, lightly damped structures in cross flow [8]. While studies on galloping and fluttering in railway systems exist, most listed studies are focused on computational fluid dynamics [9].

Some notable works on Maglev (magnetically levitated) trains that studied the beam's reaction force, the electromagnetic force, and the aeroelastic force, either individually or in combination, are listed here. Wu et. al. [10] studied the suspension stability of a Maglev vehicle under steady aerodynamic loading which consists of lift and pitching moments. Wang et. al. [11] considered time delay speed feedback effects on the linear stability and dynamic behaviour of the Maglev system and Zhang et al. extended the work by measuring time delays from two sources, the gap sensor and accelerometer [12]. One of the early works by Cai et. al [13] showed the stability of Maglev systems based on experimental data, scoping calculations and simple mathematical models. Schneider et. al. [14] introduced model of a detailed rigid multibody Maglev vehicle with three sections moving along an infinite periodically pillared elastic guideway combining the two-dimensional heave-pitch motion of the vehicle and the elastic bending of the guideway elements. A notable study in this area by Yau [15] developed a computational framework to analyse wind effects on a Maglev vehicle over flexible guideways, using PID control and proposed a PID+LQR controller to enhance ride comfort. A detailed review of the dynamic stability of repulsive-force Maglev systems can be found in [16]. The combination of the two potentially destabilizing forces, the beam's reaction force (i.e., wave-induced instability) and the electromagnetic force, was conducted by Faragau et al. [17], who determined how stability regions for control parameters are affected by the vehicle's velocity. They also identified limit cycles in a specific region of the control parameter plane.

The present work conducts a detailed study of the interaction between the electromagnetic and aeroelastic instability mechanisms in the context of a Hyperloop vehicle. One of the key findings of the present paper is the suppression of parametric resonance through the use of a (added) linear parametric force. Several notable studies have explored the suppression of parametric resonance. Yabuno et al. [18] examined electromagnetic levitation under base excitation and achieved parametric resonance suppression using a pendulum with a controller, marking one of the early contributions to this field. The suppression of parametric resonance was achieved as the nonlinear action of the pendulum on the main system counteracting the effect of the resonant parametric excitation. However, in the current work, suppression is achieved via a different state-dependent force, with the controller playing an indirect role. Inoue et al. explored the same system with excitation on the mass, employing linear PD control [19]. Another well-known approach for the suppression of parametric resonance is the redirection of energy introduced into the system to nonlinear energy sinks (NES) [20]; a detailed investigation into various applications of NES can be found in [21]. Passive nonlinear vibro-impact attachments can also be employed [22]. Recently, Pumhössel introduced a novel concept for suppressing parametric resonance through the use of state-dependent impulses [23,24]. In [25], a non-linear control law utilizing a two-frequency signal, which interacts with the parametrically excited mode through a sub-combination resonance, is implemented to suppress the parametric resonance.

The current paper can be divided into two major sections; in the first part, the interaction of the electromagnetic and aeroelastic instability mechanisms is studied for constant coefficients, and in the second part, the interaction is studied for periodically varying coefficients. The paper is structured as follows. Section 1 provides an introduction, followed by a problem statement in Section 2. Section 3 presents the stability analysis for constant coefficient values, while Section 4 explores periodically varying coefficients, highlighting the phenomenon of parametric resonance and its suppression. Section 5 concludes the study.

## 2   Problem statement

Figure 1 illustrates the considered model, representing a simplified model of a Hyperloop vehicle of mass $m$ suspended from a fixed support through the electromagnetic force, $F_e$. The support may undergo an oscillation $A\cos(\Omega t)$ with amplitude $A$, which renders the (linearized, as shown below) electromagnetic force a parametric one (i.e., it is proportional to the response variables and has time-periodic coefficients). The mass is also subject to the aeroelastic force, $F_a$, which represents an additional instability mechanism. The aeroelastic force has a part with a constant coefficient, and a part with a time-periodic coefficient can be added to it; we refer to the latter as the parametric aeroelastic force. This modulation could be achieved using flaps or teeth as used in commercial aircrafts. For example, it is well established that leading-edge modifications of an aerofoil structure, such as a drooped leading edge (using flaps), can significantly alter the aeroelastic coefficients (see Figs. 14-19 in [26]). That means, controlling the drooping of the leading edge controls the aeroelastic force. Another option is the introduction of teeth on the aerofoil surface, inspired by the denticles that cover shark skin [27]. A drooped leading edge provides greater control over the magnitude of aerodynamic coefficients; however, frequency control is limited due to the presence of relatively large inertial structures. In contrast, for the latter case, the profile and orientation of these small teeth can be easily adjusted to meet specific frequency requirements. It should be noted, however, that the specific design to achieve the parametric aeroelastic force is not subject of the present paper.

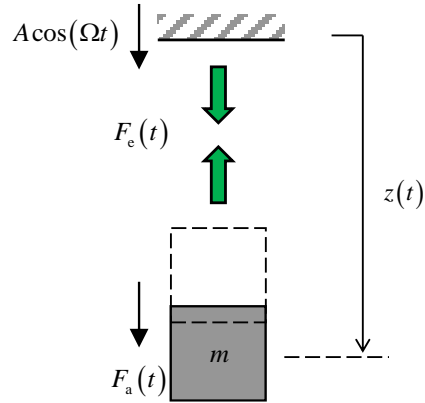

Figure 1. Model of electromagnetically suspended mass subject to air flow. Here, $F_e(t)$ is the electromagnetic force and $F_a(t)$ is the aeroelastic force.

We split the analysis into two parts: the case with coefficients constant (no parametric excitation) and the case with periodically varying coefficients (parametric-excitation, due to oscillations in the support and in the aeroelastic coefficient). We will discuss the parametric-excitation case in detail in Section 4, for which we need different equations of motion (EoMs). For now, we start from the first case and the following EoMs are considered; the first is Newton's second law, and the second is the equation for the electric current, which includes voltage control (i.e., PD control):

$$m\ddot{z}(t) = -F_e(t) + mg + F_a(t) = -C\frac{I^2(t)}{z^2(t)} + mg + \mu \dot{z}(t) \qquad (1)$$

$$\mu = -\frac{1}{2}\rho A_c \left(\frac{\partial C_z}{\partial \alpha}\right)_{\alpha=0} V \qquad (2)$$

$$\dot{I}(t) + \frac{z(t)}{2C}\left(R - 2C\frac{\dot{z}(t)}{z^2(t)}\right)I = \frac{z(t)}{2C}\left(u_0 + K_\mathrm{p}(z(t)-z_0) + K_\mathrm{d}\dot{z}(t)\right); \quad R = \frac{u_0}{I_0} \qquad (3)$$

The system operates within the gravitational field, experiencing downward acceleration $g$ due to gravity. The desired fixed gap between the vehicle and support, denoted by $z_0$, corresponds to one of the fixed points with respective steady-state voltage $u_0$ and current $I_0$ (see Section 3). The electromagnetic force $F_\mathrm{e}(t)$ between the support and vehicle depends on the displacement $z(t)$ and current $I(t)$ variables. The voltage controls the electromagnet (i.e., the current) to maintain the gap as constant as possible, with control parameters $K_\mathrm{p}$ and $K_\mathrm{d}$. $C$ is a constant determined by electromagnet properties. The destabilizing term $\mu\dot{z}$ in Eq. (1) represents the aeroelastic force with constant coefficient [10], with $\mu$ being the product of a number of constants. Here, $\alpha$ denotes the relative angle between horizontal wind velocity (with magnitude $V$) and vertical component of the vehicle velocity $\dot{z}$, $\rho$ is the air density, $A_\mathrm{c}$ is the vehicle's cross-sectional area experienced by the wind, and $C_z(\alpha) = C_\mathrm{L}(\alpha) + C_\mathrm{D}(\alpha)$, where $C_\mathrm{L}(\alpha)$ denotes the lift coefficient and $C_\mathrm{D}(\alpha)$ the drag coefficient. For galloping [5], a straightforward derivation of the destabilizing term $\mu\dot{z}$ is given in [6]. While $\mu$ is not a constant in real cases, the maximum oscillation amplitude of the vehicle, typically in the millimetre range, justifies the assumption due to minimal angle change over time.

## 3  Stability analysis for constant coefficients (no parametric excitation)

To understand the influence of parametric excitation on the stability of the equilibrium point (i.e., the shape of the stable zones), we first analyse the stability of the (relevant) equilibrium point without parametric excitation. Thus, we analyse the stability of the linearized system and we explore limit cycles for the case of $A = 0$ and the aeroelastic force having a constant coefficient.

### 3.1  Linear stability analysis

This section undertakes linear stability analysis. The approach involves linearizing Eqs. (1) and (3), and deriving eigenvalues of the Jacobian matrix obtained from the linearized equations set at the desired fixed point. Initially, fixed points are determined by considering equilibrium or steady states. The equilibrium states, where all time derivatives are zero, are described by the following set of algebraic equations (obtained from Eqs. (1) and (3)):

$$\begin{aligned} C\frac{I_0^2}{z_0^2} &= mg \\ \frac{z}{2C}\frac{u_0}{I_0}I &= \frac{z}{2C}\left(u_0 + K_\mathrm{p}(z-z_0)\right) \end{aligned} \qquad (4)$$

Solving Eq. (4) results in two fixed points:

$$\begin{aligned} z_\mathrm{ss}^\mathrm{a} &= z_0; \quad I_\mathrm{ss}^\mathrm{a} = I_0 \\ z_\mathrm{ss}^\mathrm{b} &= \frac{-z_0(u_0 - K_\mathrm{p}z_0)}{u_0 + K_\mathrm{p}z_0}; \quad I_\mathrm{ss}^\mathrm{b} = \frac{I_0(u_0 - K_\mathrm{p}z_0)}{u_0 + K_\mathrm{p}z_0} \end{aligned} \qquad (5)$$

For the second fixed point, either $z_{ss}^b$ or $I_{ss}^b$ must be negative, rendering it a nonphysical equilibrium point, especially for systems like Hyperloop. Hence, for subsequent analyses, only the fixed point $z_{ss}^a = z_0$; $I_{ss}^a = I_0$ is considered (unless mentioned otherwise).

The next step is to derive the linearized equations. Assuming perturbations around the variables as $z = z_0 + \Delta_{tr}(t)$ and $I(t) = I_0 + I_{tr}(t)$ (the subscript "tr" denotes transient), and applying Taylor series expansions up to and including first order yields

$$m\ddot{\Delta}_{tr} = -2C\frac{I_0}{z_0^2}I_{tr} + 2C\frac{I_0^2}{z_0^3}z_{tr} + \mu\dot{\Delta}_{tr}$$
$$\dot{I}_{tr} = -\frac{z_0 R}{2C}I_{tr} + \frac{z_0 K_p}{2C}\Delta_{tr} + \frac{(K_d z_0^2 + 2CI_0)}{2z_0 C}\dot{\Delta}_{tr} \tag{6}$$

The Jacobian of Eq. (6) at the fixed point-a is defined when Eq. (6) is written in state-space form:

$$\frac{d}{dt}\begin{pmatrix}\Delta_{tr}(t)\\ \dot{\Delta}_{tr}(t)\\ I_{tr}(t)\end{pmatrix} = \begin{pmatrix} 0 & 1 & 0 \\ \dfrac{2CI_0^2}{mz_0^3} & \dfrac{\mu}{m} & -\dfrac{2CI_0}{mz_0^2} \\ \dfrac{K_p z_0}{2C} & \dfrac{K_d z_0}{2C} + \dfrac{I_0}{z_0} & -\dfrac{u_0 z_0}{2CI_0} \end{pmatrix}\begin{pmatrix}\Delta_{tr}(t)\\ \dot{\Delta}_{tr}(t)\\ I_{tr}(t)\end{pmatrix} \tag{7}$$

The characteristic polynomial of the Jacobian given in Eq. (7) is

$$\lambda^3 - \left(\frac{u_0 z_0}{2CI_0} - \frac{\mu}{m}\right)\lambda^2 - \left(\frac{K_d I_0}{mz_0} - \frac{\mu u_0 z_0}{2CI_0 m}\right)\lambda + \frac{I_0(u_0 - K_p z_0)}{mz_0^2} = 0 \tag{8}$$

The eigenvalues for each of the fixed points are shown in Figure 2 (i.e., also for fixed point-b). Stability transitions can be obtained from the zero crossings of the real parts of the eigenvalues.

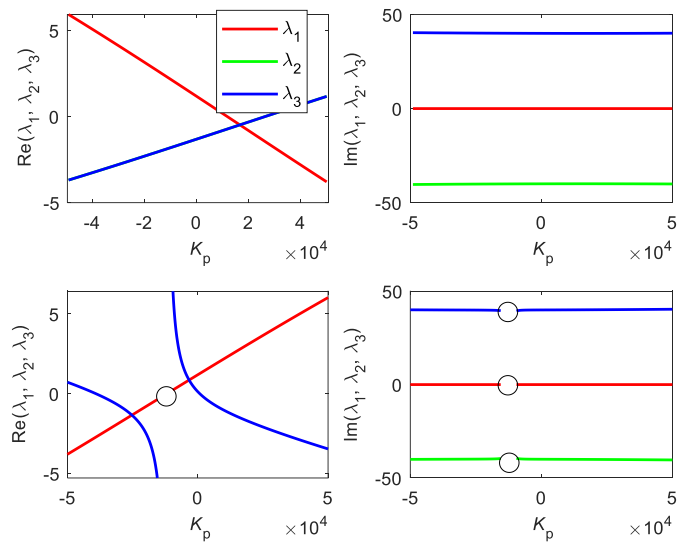

Figure 2: The eigenvalues for each of the fixed points are shown here, the upper and lower panels represent first and second fixed points, respectively. The small circles in the lower panels represent singularities. Here, $K_\text{d} = 10000\,(\text{Vs/m})$, $C = 0.05\,(\text{Nm}^2/\text{A}^2)$, $z_0 = 0.015\,(\text{m})$, $m = 7650\,(\text{kg})$.

Utilizing properties of cubic polynomials, the stability boundaries related to the first equilibrium point can be determined. The discriminant of the polynomial suggests that the roots contain one real and two complex conjugates (not shown here). A stability transition requires at least one eigenvalue's real part to be zero (sign change), suggesting two possibilities: the real part of the complex conjugates is zero, or the real root is zero.

In the first scenario, for a polynomial $\lambda^3 + a\lambda^2 + b\lambda + c = 0$ to have one real root and two purely imaginary roots, the relation $ab = -c$ is required, resulting in the first stability transition which is a straight line in the $K_\text{p}$-$K_\text{d}$ plane (see Figure 3):

$$K_\text{p} = \frac{u_0}{z_0} + \frac{2CI_0\mu^2 u_0 z_0^3 - m\mu u_0^2 z_0^4}{4C^2 I_0^3 m z_0} + \frac{2CI_0^2 m u_0 z_0^2 - 4C^2 I_0^3 \mu\, z_0}{4C^2 I_0^3 m z_0} K_\text{d} \tag{9}$$

In the second scenario, the value $c$ will be zero since there will be only two non-zero roots, leading to the following condition, which is a vertical line in Figure 3:

$$K_\text{p} = \frac{u_0}{z_0} \tag{10}$$

The requirement for unconditional instability can be determined when the slope of Eq. (9) approaches infinity and coincides with the left vertical line in Figure 3:

$$\mu = \frac{m u_0 z_0}{2CI_0} \tag{11}$$

In the limit where there is no influence of aeroelastic force ($\mu = 0$), the stability boundary (see Eq. (9)) reduces to:

$$K_\text{p} = \frac{u_0}{z_0} + \frac{u_0 z_0}{2CI_0} K_\text{d} \tag{12}$$

and the natural frequency of the system at the right boundary, obtained from the purely imaginary eigenvalues, is given (for later use) as,

$$\omega_0 = \sqrt{\frac{K_\text{d} I_0}{m z_0}} \tag{13}$$

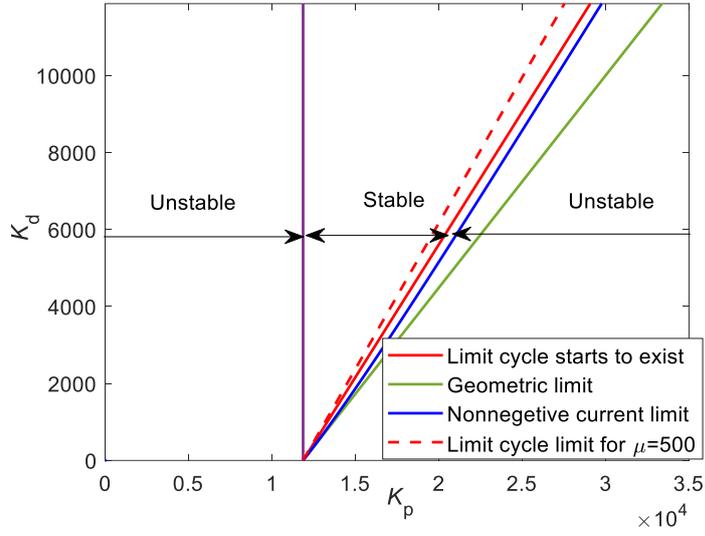

Figure 3: Stability regions for the first fixed point in the $K_p$ - $K_d$ plane. Here,
$$C = 0.05\left(\text{Nm}^2/\text{A}^2\right),\ z_0 = 0.015\,(\text{m}),\ m = 7650\,(\text{kg})$$

## 3.2  Determination of limit cycle for the case $\mu = 0$

The analysis now shifts its focus to the nonlinear dynamics aspect. It is evident from the stability analysis provided earlier that when the real part of the complex conjugate roots equals zero, the corresponding solution is a harmonic motion, typically indicating the presence of a limit cycle or periodic solution in the vicinity. This bears resemblance to the supercritical Hopf bifurcation, albeit typically defined for single-degree-of-freedom systems. We employ the harmonic balance method [28] for the determination of the limit cycle. Notably, the harmonic balance analysis differs between cases with and without aeroelastic force, hence treated separately in two sections.

Here, we delve into the scenario without aeroelastic force. Upon careful examination of the EoMs given in Eqs. (1) and (3) for $\mu = 0$, two key observations emerge. First, the system is autonomous, allowing us to arbitrarily choose the time origin as follows: $z(0) = 0$. Second, there exists no first-order time derivative for either variable $z(t)$ or $I(t)$ in Eq. (1), ensuring a zero phase shift between them. For instance, selecting $z = z_0 + a\cos(\omega t)$ and $I = I_0 + b\cos(\omega t) + c\sin(\omega t)$ would render $c$ as zero (shown in Eq. (32) when $\mu = 0$), as there would only be one sine term upon substituting these assumptions into Eq. (1). However, the scenario changes entirely when $\mu \neq 0$, introducing a slightly more intricate derivation process, elaborated upon in the subsequent section.

Utilizing harmonic balance, we examine the presence of a limit cycle, truncating after the first harmonic. Let us assume that,

$$z = z_0 + a\cos(\omega t) \tag{14}$$

$$I = I_0 + b\cos(\omega t) \tag{15}$$

Substituting Eqs. (14) and (15) into (1) and (3) and rearranging results in one equation of the following form:

$$M_0 + M_1 \sin(\omega t) + N_1 \cos(\omega t) = 0 \tag{16}$$

Equating the coefficients $M_1$ and $N_1$ of each harmonic to zero gives a system of three algebraic equations in terms of the unknowns $a, b, \omega$:

$$8aCI_0^2 - 8bCI_0z_0 + 3a^3m\omega^2 z_0 + 4am\omega^2 z_0^3 = 0$$
$$3a^3I_0K_p - 3a^2bu_0 + 4aI_0K_p z_0^2 - 4bu_0 z_0^2 = 0 \quad (17)$$
$$8aCI_0 + a^3K_d - 8bCz_0 + 4aK_d z_0^2 = 0$$

This gives

$$a = \frac{2\sqrt{2CI_0K_p z_0 - 2CI_0 u_0 - K_d u_0 z_0^2}}{\sqrt{K_d}\sqrt{u_0}} \quad (18)$$

$$b = \frac{2I_0 K_p \sqrt{-2CI_0 u_0 + 2CI_0 K_p z_0 - K_d u_0 z_0^2}}{\sqrt{K_d}\, u_0^{3/2}} \quad (19)$$

$$\omega = \frac{I_0 \sqrt{K_d}\sqrt{C(K_p z_0 - u_0)}}{\sqrt{mz_0}\sqrt{3CI_0 K_p z_0 - 3CI_0 u_0 - K_d u_0 z_0^2}} \quad (20)$$

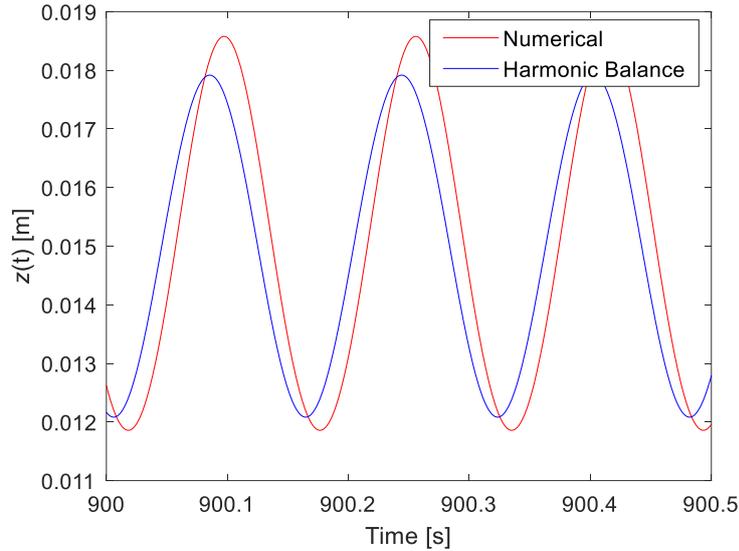

Figure 4. Comparison of numerical integration results and harmonic balance prediction for a limit cycle for $\mu = 0$. Here, $K_p = 27000\,(\text{V/m})$, $K_d = 10000\,(\text{Vs/m})$, $C = 0.05\,(\text{Nm}^2/\text{A}^2)$, $z_0 = 0.015\,(\text{m})$, $m = 7650\,(\text{kg})$.

Eqs. (18)-(20) completes the identification of limit cycle for $\mu = 0$.

Results obtained from numerical integration and those from harmonic balance are compared in Fig. 4. A small difference can be observed, which could have been anticipated due to the neglection of the higher harmonics in the analytical result. For the existence of a limit cycle, $a, b, \omega \geq 0$ must hold true; based on that condition, exactly the right boundary of the stable domain (Eq. (12)) is obtained. In other words, the limit cycle is born the moment that the fixed point becomes unstable (depicted by the inclined red line in Fig. 3).

From Eq. (18), it is clear that $a$ is dependent on many parameters and, at the same time, the oscillations are limited up to $a \leq z_0$; otherwise, the mass hits the boundary. The geometrical constraint $a = z_0$ leads to the following line in the $K_p$ - $K_d$ plane:

$$K_p = \frac{CI_0^2 K_d u_0 - 3CI_0 m u_0 z_0^3 - K_d m u_0 z_0^5}{CI_0^2 K_d z_0 - 3CI_0 m z_0^4} \tag{21}$$

Beyond this line, the limit cycle can no longer exist. Similarly, there is physical constraint on the current oscillation; the current is assumed as nonnegative (for real systems like Hyperloop), i.e. $b \leq I_0$. The limit $b = I_0$ leads to

$$K_d = \frac{8CI_0 K_p^2 \left(K_p z_0 - u_0\right)}{u_0^3 + 4K_p^2 u_0 z_0^2} \tag{22}$$

Furthermore, it is noteworthy that when $b \to I_0$, the numerical integration can exhibit inaccuracies as the current variable approaches zero at many instances, potentially leading to substantial asymmetry in $z(t)$ or $I(t)$ around the respective mean position.

While not exploring the second fixed point, it is acknowledged that for the second fixed point, a distinct stable region emerges (as well as a region with a limit cycle) for negative $K_p$ and $K_d$. We refrain from delving into these details as they are not of immediate physical relevance. The stable region is evident from the real and imaginary eigenvalue plots provided in Fig. 2.

### 3.3 Simplification of EoMs

When including the aeroelastic force, obtaining an analytical expression for the limit cycle becomes cumbersome due to additional terms involved in the calculation (due to the aeroelastic term in the EoMs), leading to a phase shift. It is beneficial to simplify the EoMs (without compromising accuracy). One of the EoMs provided in Eq. (1) or Eq. (3) should be linearized, but the crucial question then is: which EoM should be linearized?

Let us again consider the case $\mu = 0$. From the perspective of harmonic balance, it becomes evident that linearizing Eq. (3) is not advisable. The reason is that we require three equations to solve for the three variables $a, b$ and $\omega$, and at least two of these equations must be nonlinear to obtain nonzero solutions. Eq. (1) provides only one algebraic equation as there are no first-order derivatives for $\mu = 0$, and it is always homogeneous. Linearizing Eq. (3) would yield two linear algebraic equations, which are homogeneous too, resulting in only zero solutions. Thus, the only viable option is to linearize Eq. (1) while keeping Eq. (3) as nonlinear. The simplification thus proceeds as follows. After linearizing Eq. (1), we obtain

$$m\ddot{z}(t) + \frac{2CI_0 I(t)}{z_0^2} - \frac{2CI_0^2 z(t)}{z_0^3} - \mu \dot{z}(t) = 0 \tag{23}$$

We can then eliminate $I(t)$ by solving for $I(t)$ from Eq. (23) and substitute it into Eq. (3). The simplified, single-variable EoM appears as follows:

$$-2CI_0^2\left(u_0 - K_p z_0\right)z^3(t) + 2CI_0 z_0^3 \dot{z}(t)\left(\mu \dot{z}(t) - m\ddot{z}(t)\right) +$$
$$+ z_0 z^2(t)\left(2CI_0^2\left(u_0 - K_p z_0\right) + \left(2CI_0^2 K_d - \mu u_0 z_0^2\right)\dot{z}(t) + m u_0 z_0^2 \ddot{z}(t)\right) + \quad (24)$$
$$+ 2CI_0 z_0^3 z(t)\left(-\mu \ddot{z}(t) + m\dddot{z}(t)\right) = 0$$

It can be verified that the simplified EoM provided in Eq. (24) yields an accurate prediction of the limit cycle (close to 1% error) compared to that obtained from the full nonlinear set of equations Eqs. (1) and (3). Even though the EoM is related to a 1.5DOF system, its behaviour demonstrates an analogy with the supercritical Hopf bifurcation, in the following manner; linear stability analysis indicates that fixed point-a transitions from stable to unstable at the red boundary in Fig. 2 with an increase in $K_p$ for constant $K_d$. Harmonic balance analysis reveals the existence of a limit cycle on the unstable side of the fixed point, characteristic of a supercritical Hopf bifurcation, now obtained from a *single-variable* EoMs. The stability of the limit cycle is confirmed using Floquet analysis numerically [29].

### 3.4 Determination of limit cycle for the case $\mu \neq 0$

In this section, the harmonic balance method is used to determine the limit cycle for the equilibrium point of the system that is subject to the aeroelastic force. As mentioned before, there will be a phase shift between variables. We use the simplified EoM (Eq. 24) with aeroelastic term (the method is similar to that in Section 3.2). The major advantage of using simplified EoM given in Eq. (24) is that, if we assume $z = z_0 + a\cos(\omega t)$ and $I = I_0 + b\cos(\omega t) + c\sin(\omega t)$, the simplified EoM allows the calculation of $a$ and $\omega$ irrespective of $b$ and $c$ since the EoM is independent of $I(t)$; this gives an elegant solution procedure. Substituting $z(t) = z_0 + a\cos(\omega t)$ into Eq. (24) gives $a$ and $\omega$ as follows:

$$\omega = \sqrt{\frac{\sqrt{Q_1^2 - Q_2}}{6CI_0 m^2 u_0 z_0^4} + Q_3} \quad (25)$$

$$a = \frac{1}{\sqrt{2CI_0^2 K_d - \mu u_0 z_0^2}}\left(\sqrt{\frac{4\sqrt{Q_1^2 - Q_2}}{3m u_0 z_0} + Q_4}\right) \quad (26)$$

where

$$Q_1 = -2C^2 I_0^3 K_d \mu z_0^2 - 6C^2 I_0^3 K_p m z_0^2 + 6C^2 I_0^3 m u_0 z_0 - \\ - 2CI_0^2 K_d m u_0 z_0^3 + CI_0 \mu^2 u_0 z_0^4 + m\mu u_0^2 z_0^5 \quad (27)$$

$$Q_2 = 12CI_0 m^2 u_0 z_0^4 \left(4C^2 I_0^4 K_d K_p z_0 - 4C^2 I_0^4 K_d u_0 - 2CI_0^2 K_p \mu u_0 z_0^3 + 2CI_0^2 \mu u_0^2 z_0^2\right) \quad (28)$$

$$Q_3 = \frac{CI_0^2 K_d \mu}{3m^2 u_0 z_0^2} + \frac{CI_0^2 K_p}{m u_0 z_0^2} - \frac{CI_0^2}{m z_0^3} - \frac{\mu u_0 z_0}{6CI_0 m} + \frac{I_0 K_d}{3m z_0} - \frac{\mu^2}{6m^2} \quad (29)$$

$$Q_4 = \frac{8C^2 I_0^3 K_d \mu z_0}{3m u_0} + \frac{8C^2 I_0^3 K_p z_0}{u_0} - 8C^2 I_0^3 - \frac{16}{3}CI_0^2 K_d z_0^2 - \frac{4CI_0 \mu^2 z_0^3}{3m} + \frac{8}{3}\mu u_0 z_0^4 \quad (30)$$

Now that the variables $a$ and $\omega$ are known, an expression for $I(t)$ can be derived in a straightforward manner. Applying harmonic balance to Eq. (23) by substituting $I(t) = I_0 + b\cos(\omega t) + c\sin(\omega t)$ as well as the expression for $z(t)$ gives,

$$b = \frac{2aCI_0^2 + am\omega^2 z_0^3}{2CI_0 z_0}; \quad c = -\frac{a\mu\omega z_0^2}{2CI_0} \tag{31}$$

Note that $c = 0$ when $\mu = 0$, as discussed in Section 2.2.

## 4  Stability analysis for harmonically varying coefficients

In this section, we introduce parametric excitation through the periodic variation of the coefficients of both the (linearized) electromagnetic and aeroelastic forces (the former is a result of the base excitation applied). First, we consider the simple case where only the coefficients of the electromagnetic force are time dependent (we refer to it as the parametric electromagnetic force), and then we add a part to the aeroelastic force which has a harmonically varying coefficient (we refer to the added force as the parametric aeroelastic force). Linearized EoMs around the time-varying equilibrium/steady state are used for the stability analysis. For numerical calculations, we employ the Floquet method [29], while the analytical approach to find the stability boundary specifically related to parametric resonance utilizes Hill's method; for the part away from the zone of parametric resonance we also use a Hill's type method. Interestingly, the parametric resonance is characterized by an elliptical region, and we provide a simple expression to describe the ellipse. When the parametric aeroelastic force is added, the expression for the instability boundary reveals a complicated and nontrivial dependence of the size of the ellipse on the phase difference. Energy analysis uncovers that the parametric aeroelastic force can play a dual role in determining the severity of the net parametric resonance, caused by the interplay of it with the other state-dependent force, that is, the parametric electromagnetic force.

Here, we introduce two major additions to the EoMs defined in Eqs. (1)-(3); an excitation is applied to the rigid base and an added parametric aeroelastic force, is considered:

$$\Delta_{ss}(t) = z_0 - A\cos(\Omega t); \quad \mu(t) = \mu_0 + \mu_1 \cos(\Omega t - \phi) \tag{32}$$

Here, $\Delta_{ss}$ is the new, steady-state air gap which is time varying due to the base excitation, and $z_0$ is chosen to be the same as for the unforced system. The two cases described above are considered one by one in the following sections.

### 4.1  Parametric resonance for the case $\mu_0 = 0$, $\mu_1 = 0$

Let us consider the first case, in which the base excitation is applied but there is no aeroelastic force. We define the following perturbations to linearize the system

$$\begin{aligned} z(t) &= z_0 + \Delta_{tr}(t) \\ I(t) &= I_{ss}(t) + I_{tr}(t) \end{aligned} \tag{33}$$

After substitution of Eq. (33) into Eqs. (1)-(3), and elimination of nonlinear terms, the linearized EoMs read as follows:

$$m\frac{d^2\Delta_{tr}}{dt^2} = -\frac{2CI_{ss}^2}{\Delta_{ss}^3}\left(\frac{\Delta_{ss}}{I_{ss}}I_{tr} - \Delta_{tr}\right) \tag{34}$$

$$\frac{dI_{tr}}{dt} + \frac{R\Delta_{ss}^2 - 2C\dot{\Delta}_{ss}}{2C\Delta_{ss}}I_{tr} = \left(\frac{K_p}{2C}\Delta_{ss} - \frac{\dot{\Delta}_{ss}I_{ss}}{\Delta_{ss}^2}\right)\Delta_{tr} + \left(\frac{K_d}{2C} + \frac{I_{ss}}{\Delta_{ss}^2}\right)\Delta_{ss}\dot{\Delta}_{tr} \tag{35}$$

In Eqs. (34) and (35), the steady-state current $I_{ss}$ is defined as,

$$I_{ss}(t) = \sqrt{\frac{mg}{C}}\Delta_{ss} \tag{36}$$

It is possible to eliminate $I_{tr}$ from Eqs. (34) and (35) and obtain a very simplified single EoM:

$$2\left(CK_p\sqrt{\frac{gm}{C}} - gmR\right)\Delta_{tr} + 2CK_d\sqrt{\frac{gm}{C}}\dot{\Delta}_{tr} + z_0 m R\ddot{\Delta}_{tr} + 2Cm\ddot{\Delta}_{tr} - A\cos(\Omega t)mR\ddot{\Delta}_{tr} = 0 \tag{37}$$

Eq. (38) is the starting point for the derivation using Hill's determinant method, presented below, which aims at determining the stability boundary, and also for the numerical validation using Floquet theory [29] (see Figures 5 and 6). For the Hill's determinant method, let us represent the solution by a complex Fourier series:

$$\Delta_{tr} = \sum_{n=-\infty}^{\infty} d_n \exp(in\omega t); \; \Omega = k\omega; \; n,k \in \mathbb{Z} \tag{38}$$

Some important observations are given here. First, the term with $n=0$ gives the left, vertical boundary of stability region (see Fig. 5); taking $\Delta_{tr} = d_0$ and substituting that in Eq. (37), we obtain

$$CK_p\sqrt{\frac{gm}{C}} - gmR = 0 \Rightarrow K_p = \frac{u_0}{z_0} \tag{39}$$

The result in Eq. (39) is the vertical boundary, as also expressed in Eq. (10); clearly, the left stability boundary is the same for the unforced and forced systems.

Second, $k=1$ represents the $T_1$ parametric resonance and $k=2$ represents the $T_2$ parametric resonance. In the following derivations, we assume a first-harmonic ($|n|=1$) approximation of the Fourier series. For the situation with $k>2$, we can verify that (after substitution Eq. (38) into Eq. (37)) the term $A\cos(\Omega t)mR\ddot{\Delta}_{tr}$ does not contribute to the coefficient of the leading-order harmonic (since similar terms as $|n|=1$ can only result from $k=1,2$). In the following derivations, we only consider $T_2$ parametric resonance ($k=2$) which is the comparatively most commonly encountered one; the $T_1$ boundary is very small. Unless mentioned otherwise, from here we consider $k=2$ or $\Omega=2\omega$.

Substituting Eq. (38) in Eq. (37), extracting the Hill's matrix and equating its determinant to zero gives the following expression for the stability boundary; the coordinates of the centre are $(h_1, h_2)$ with $k_1$ as the major axis and $k_2$ is the minor axis:

$$\frac{(K_p - h_1)^2}{k_1^2} + \frac{(K_d - h_2)^2}{k_2^2} = 1 \tag{40}$$

$$h_1 = \frac{R\left(2\sqrt{Cg^3m} + \sqrt{Cgm}z_0\omega^2\right)}{2Cg}$$

$$h_2 = \sqrt{\frac{Cm}{g}}\omega^2 \tag{41}$$

$$k_1 = \omega k_2$$

$$k_2 = \sqrt{\frac{A^2mR^2\omega^2}{16Cg}}$$

From $h_2$ (i.e., the $K_d$ coordinate of the ellipse), it is possible to derive (by substituting $h_2 = K_d$ and using $I_0 = z_0\sqrt{mg/C}$ in Eq. (41) to relate $\omega = \Omega/2$ to the natural frequency $\omega_0$ of the unforced system Eq. (13)) the location of the parametric resonance ellipse on the inclined (right) stability boundary (see also Fig. 5). The $\omega_0$ of the system varies smoothly along the inclined line (Eq. (13)). Like for the classical Mathieu equation, the (first) zone of $T_2$ parametric resonance is found (i.e., the centre of the ellipse) at the point where $\omega_0/\Omega = 1/2$; the (first) $T_1$ parametric resonance zone, although very small, is found in principle found where $\omega_0/\Omega = 1$. Note that higher zones for $T_1$ and $T_2$ are not observed for the current problem.

## 4.2 Right stability boundary part unrelated to parametric resonance for the case $\mu_0 = 0$, $\mu_1 = 0$

In the previous section, the stability boundary related to $T_2$ parametric resonance has been determined. However, this is not the complete stability boundary; the system can also undergo a stability transition (i.e., become unstable) away from the elliptical boundary (which is demonstrated in Fig. 5, for example). In this section, we determine an expression for the remaining part of the right stability boundary and proof that it is the same as that of the unforced system.

As mentioned above, for $|n| = 1$ and for all the cases with $k > 2$ the term $A\cos(\Omega t)mR\ddot{\Delta}_{tr}$ does not contribute to the coefficient of the leading-order harmonic in Eq. (37). Hence, the following simple following expression is obtained:

$$K_d = \frac{2iK_p\sqrt{Cgm} - 2igmR + m\omega^2(-iRz_0 + 2C\omega)}{2\sqrt{Cgm}\omega} \tag{42}$$

Equating the imaginary part of Eq. (42) to zero gives an expression for the oscillation frequency $\omega$ which is exactly the same as the natural frequency $\omega_0$ of system without excitation (Eq. (13)). Then, substituting this expression into Eq. (42) the same straight line as given in Eq. (12) is obtained:

$$K_p = \frac{u_0}{z_0} + \frac{u_0z_0}{2CI_0}K_d \tag{43}$$

Thus, we find exactly the same inclined stability boundary as we found for the equilibrium point without excitation. However, as $k$ is an integer number in the current analysis and $\Omega$ therefore is an integer multiple of $\omega$, Eq. (43) only holds for discrete points along the straight line where $\omega/\Omega = \omega_0/\Omega = 1/k$. Furthermore, when $k > 2$, Eq. (43) is not related to a parametric-resonance type instability; by crossing the straight line, the control of the electromagnetic force becomes simply inappropriate which leads to a loss of stability.

To demonstrate that the result in Eq. (43) is also generally valid (i.e., it does describe the entire straight part of the right stability boundary), we assume the following solution:

$$\Delta_{tr} = U(t)\exp(i\omega t); \quad U(t) = U(t+T), \quad T = 2\pi/\Omega \tag{44}$$

This solution is directly based on Floquet's theorem, but it is evaluated at the stability boundary (hence, it is also a Hill's type solution); the magnitude of the Floquet multiplier therefore should be one, and hence $\omega$ should be real-valued. Representing the periodic part of the solution as a Fourier series

$$U(t) = \sum_{n=-\infty}^{\infty} U_n \exp(in\Omega t) \tag{45}$$

and incorporating only the constant, Eq. (45) can be written as

$$\Delta_{tr} = U_0 \exp(i\omega t) \tag{46}$$

Substituting Eq. (46) into Eq. (37), dividing by $\exp(i\omega t)$ (which appears in all terms) and projecting the resulting equation on the constant included in the Fourier series (i.e., integrating the equation from 0 to $T$), we obtain a homogenous equation for $U_0$. For nontrivial solutions, the coefficient of it must be zero. The thus obtained equation depends on $\omega$ and the system parameters, and appears to be exactly the same as Eq. (42). The straight line described by Eq. (43) is obtained from it in the way described right above it. However, now the frequency $\omega$ is not limited to integer fractions of $\Omega$, and therefore the result is generally valid. Hence, we conclude that the right stability boundary obtained for the system *without excitation* still describes the boundary of the system *with excitation* as long as we stay outside the regions of parametric resonance. We can verify that this remains true when the aeroelastic force (with and without time-dependent coefficients) is added.

### 4.3 Parametric resonance for the case $\mu_0 \neq 0$, $\mu_1 \neq 0$

Now, let us add the aeroelastic force to the system having a coefficient that oscillates around a nonzero constant. This force thus has a part with a constant coefficient $\mu_0$ and one with an oscillating coefficient having amplitude $\mu_1 < \mu_0$ (i.e., the aeroelastic force). The oscillating part has a phase difference $\phi$ to the base excitation. After adding the total aeroelastic force to Eq. (34), the following expression is obtained:

$$\begin{aligned}
m\frac{d^2\Delta_{tr}}{dt^2} &= -\frac{2CI_{ss}^2}{\Delta_{ss}^3}\left(\frac{\Delta_{ss}}{I_{ss}}I_{tr} - \Delta_{tr}\right) + \left(\mu_0 + \mu_1\cos(\Omega t - \phi)\right)\dot{\Delta}_{tr} \\
&= -\frac{2CI_{ss}^2}{\Delta_{ss}^3}\left(\frac{\Delta_{ss}}{I_{ss}}I_{tr} - \Delta_{tr}\right) + \left(\mu_0 + \mu_{1,C}\cos(\Omega t) + \mu_{1,S}\sin(\Omega t)\right)\dot{\Delta}_{tr}
\end{aligned} \tag{47}$$

Now, if we do the same derivations as the ones leading to the result in Eqs. (40) and (41), the ellipse properties can be found as follows:

$$h_1 = \frac{2\sqrt{Cg^3m^3}R + \sqrt{Cgm^3}Rz_0\omega^2 - 2\sqrt{C^3gm}\mu_0\omega^2}{2Cgm}$$

$$h_2 = \frac{\sqrt{Cgm}R(2z_0\mu_0 - A\mu_{1,C}) + 4\sqrt{C^3gm^3}\omega^2}{4Cgm}$$

$$k_1 = \omega k_2$$

$$k_2 = \sqrt{\frac{N_1 + N_2}{16Cgm}}$$

$$N_1 = (\mu_{1,C}^2 + \mu_{1,S}^2)(R^2z_0^2 + 4C^2\omega^2) + A^2R^2(\mu_0^2 + m^2\omega^2)$$

$$N_2 = 2AR(2C\omega(\mu_0\mu_{1,S} + m\mu_{1,C}\omega) + Rz_0(\mu_0\mu_{1,C} - m\mu_{1,S}\omega))$$

(48)

Here, note $h_1$ is independent of $\mu_{1,C}$ and $\mu_{1,S}$, and $h_2$ is independent of $\mu_{1,S}$.

Fig. 5 illustrates various scenarios of aeroelastic forcing and how the ellipse changes in size and position. The solid lines in Fig. 5 are obtained numerically using Floquet analysis, while the dashed line represents the analytical results presented in Eq. (48). The analytical and numerical results show a perfect match.

The black line in Fig. 5 corresponds to the case without aeroelastic forcing. Introducing an aeroelastic force with constant coefficient shifts the right stability boundary to the left shown by the red line, indicating that the aeroelastic force tends to destabilize the system. Then, by adding the parametric aeroelastic force alongside the component with the constant coefficient, the ellipse begins to shrink (red line to green line). As the amplitude of the coefficient of the parametric force increases, the ellipse completely disappears at some point (purple dot in Fig. 5). It can be verified that further increase in amplitude of the modulation coefficient causes the ellipse to grow again and reaches back the green line at around $\mu_{1,C} = 18000$. Note that the values of $\mu_0$ and $\mu_{1,C}$ in Fig. 5 have been selected to clearly illustrate the effects. However, for practical applications in Hyperloop, specific designs may be required to achieve those.

In Fig. 5, another interesting observation can be made by comparing three cases: the black, blue and grey lines. The black line represents the first individual case, where there is only the parametric electromagnetic forcing, due to base excitation, and no aeroelastic force ($A = 0.0142$, $\mu_0 = 0$, $\mu_{1,C} = 0$, $\mu_{1,S} = 0$). The blue line shows the second individual case, where there is only the parametric aeroelastic force and no parametric electromagnetic force ($A = 0$, $\mu_0 = 0$, $\mu_{1,C} = 8000$, $\mu_{1,S} = 0$). The parametric aeroelastic force and the parametric electromagnetic force have the same frequency and same phase. It is clear in Fig. 5 that both individual forces create parametric resonance, which implies that they add energy to the system (for specific values of the control parameters inside the corresponding ellipses). However, when they are combined ($A = 0.0142$, $\mu_0 = 0$, $\mu_{1,C} = 8000$, $\mu_{1,S} = 0$), the parametric aeroelastic force reverses its character and extracts energy from the system, resulting in an ellipse that is much smaller than in either of the individual cases (the detailed energy analysis given in Section 4.3.2).

Using Eq. (48), one can easily formulate the condition $k_1 = 0$, where the ellipse is eliminated and no parametric resonance occurs at all. We can find combinations of $\mu_{1,C}$ and $\mu_{1,S}$ to guarantee $k_1 = 0$. Here, we take $\mu_{1,S} = 0$; the following specific expression for $\mu_{1,C}$ is then found:

$$\mu_{1,C} = \mu_{opt} = \frac{AR^2 z_0 \mu_0 + 2ACmR\omega^2 - \sqrt{-A^2 m^2 R^4 z_0 \omega^2 + 4A^2 CmR^3 z_0 \mu_0 \omega^2 - 4A^2 C^2 R^2 \mu_0^2 \omega^2}}{R^2 z_0^2 + 4C^2 \omega^2} \qquad (49)$$

At this specific value of $\mu_{1,C}$, as given in Eq. (49), the ellipse is completely suppressed as shown as a purple dot in Fig. 5. The result given in Eq. (49) is complex, which is perhaps counterintuitive, but the imaginary part can be verified to be small. The result being complex is deemed to originate from the first-harmonic approximation of Eq. (38).

The current findings (i.e., the elimination of the ellipse) demonstrate that we can effectively suppress the parametric resonance induced by the parametric electromagnetic force, which arises from the base excitation, by introducing another parametric force (the parametric aeroelastic force) with the same frequency as the parametric electromagnetic force. It is evident that the parametric aeroelastic force can exhibit either stabilizing or destabilizing behaviour, depending on the specific amplitude of the time-varying aeroelastic coefficient. However, achieving optimal suppression of parametric resonance requires a specific combination of $\mu_1$ and $\phi$, which will be explored in the next section.

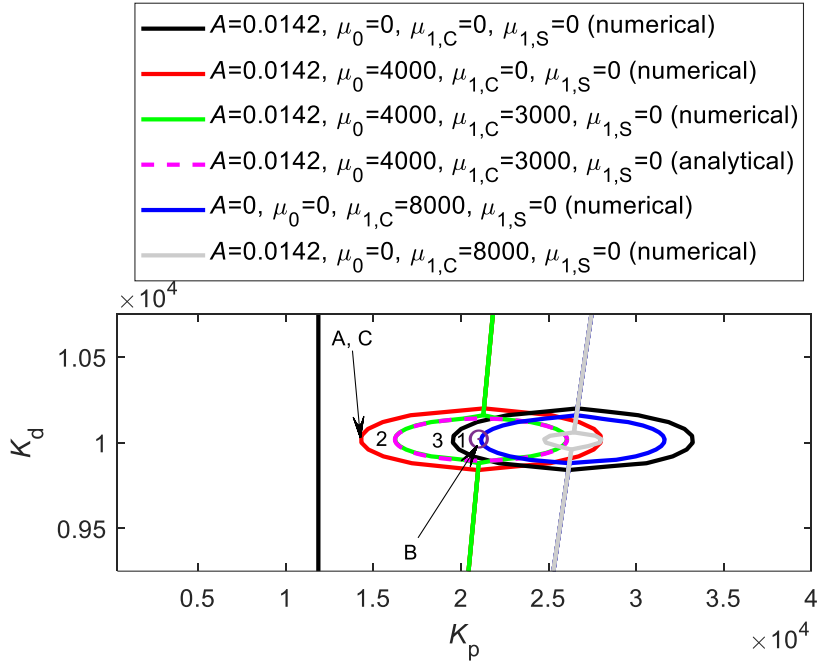

Figure 5. The elliptical region for different cases of $A$, $\mu_0$, $\mu_{1,C}$ and $\mu_{1,S}$ are shown. Solid lines show numerical results based on Floquet theory and the dashed line shows the analytical results using Eq. (48). The purple dot at point B represents the condition $k_1 = 0$ or the optimum situation. Here, $C = 0.05 \left( \text{Nm}^2/\text{A}^2 \right)$, $z_0 = 0.015 (\text{m})$, $m = 7650 (\text{kg})$, $R = 9.71 (\text{Ohm})$, $\omega = 40 (\text{rad/s})$

### 4.3.1 Effect of phase difference

From Eq. (47) we know that $\mu_{1,C} = \mu_1 \cos\phi$, $\mu_{1,S} = \mu_1 \sin\phi$ and $\mu_1 = \sqrt{\mu_{1,C}^2 + \mu_{1,S}^2}$. To explore the optimum combination of $\mu_1$ and $\phi$, we set the following condition:

$$\mu_{1,S} = \sqrt{\mu_{opt}^2 - \mu_{1,C}^2} \qquad (50)$$

Substituting Eq. (49) and Eq. (50) into Eq. (48) (i.e., the expression for $k_1$) gives a relation between the major axis $k_1$ and the phase difference through $\mu_{1,S}$:

$$k_1 = \frac{1}{4}\sqrt{\frac{AR\omega^2(M_1 + M_2 + M_3 - M_4)}{Cgm}}$$

$$M_1 = AR(\mu_0^2 + m^2\omega^2)$$

$$M_2 = \frac{AR(Rz_0\mu_0 + 2Cm\omega^2 - i\omega(mRz_0 - 2C\mu_0))^2}{R^2z_0^2 + 4C^2\omega^2}$$

$$M_3 = 2Rz_0\left(m\mu_{1,S}\omega - \mu_0\sqrt{-\mu_{1,S}^2 + \frac{A^2R^2(Rz_0\mu_0 + 2Cm\omega^2 - i\omega(mRz_0 - 2C\mu_0))^2}{(R^2z_0^2 + 4C^2\omega^2)^2}}\right)$$

$$M_4 = 4C\omega\left(\mu_0\mu_{1,S} + m\omega\sqrt{-\mu_{1,S}^2 + \frac{A^2R^2(Rz_0\mu_0 + 2Cm\omega^2 - i\omega(mRz_0 - 2C\mu_0))^2}{(R^2z_0^2 + 4C^2\omega^2)^2}}\right)$$

(51)

Like in Eq. (49), the result given in Eq. (51) is complex, with a small imaginary part; the result being complex is again deemed to originate from the first-harmonic approximation of Eq. (38).

The strong dependence of the effectiveness of the suppression on the phase shift is shown in Fig. 6. Hence, for the elimination of the parametric resonance the phase must be properly controlled. In Fig. 6, it is also interesting to note that when $\phi$ is slightly negative, the parametric aeroelastic force is still very effective in suppressing the parametric resonance. However, when $\phi$ is slightly positive, the parametric aeroelastic force rapidly loses the effectiveness. The dashed line in the inset of Fig. 6 shows the length of the major axis when there is no parametric aeroelastic force. Clearly, below the dashed line, the interaction between parametric electromagnetic and aeroelastic forces causes the ellipse to shrink, expanding the stable domain. Conversely, above the dashed line, the ellipse enlarges, reducing the stable domain.

In the next section the physical reasoning of the elimination of the parametric resonance is explored using energy analysis.

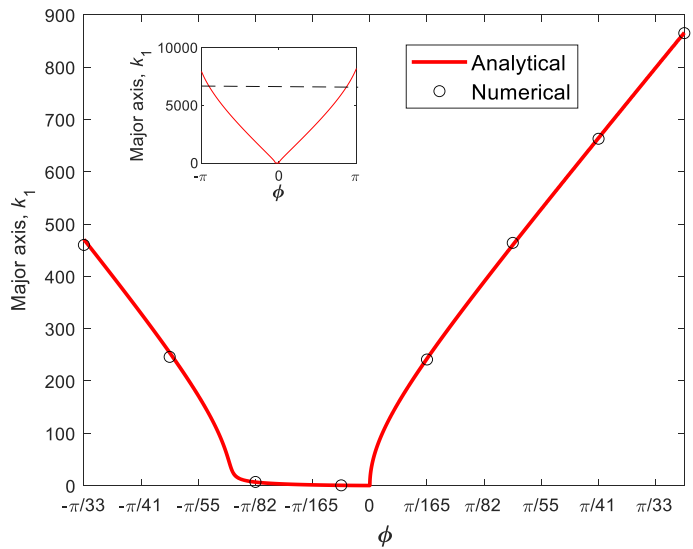

Figure 6. The dependence of the size of the ellipse related to parametric resonance on the phase difference is shown. The numerical results are obtained using Floquet analysis and the analytical results are obtained using Eq. (51). In the inset, a larger range of $\phi$ is shown. The dashed line in the inset shows the major axis ($k_1$) when there is no parametric aeroelastic force. Here, $K_p = 27000\,(\text{V/m})$, $K_d = 10000\,(\text{Vs/m})$, $C = 0.05\,(\text{Nm}^2/\text{A}^2)$, $z_0 = 0.015\,(\text{m})$, $m = 7650\,(\text{kg})$, $R = 9.71\,(\text{Ohm})$, $A = 0.0142\,(\text{m})$, $\omega = 40\,(\text{rad/s})$, $\mu_0 = 4000\,(\text{Ns/m})$

### *4.3.2 Energy analysis*

In this section, an energy analysis for the system represented by Eq. (47) is conducted. The anomalous behaviour of the parametric aeroelastic force compared to the aeroelastic force with constant coefficient is investigated in specific ranges of $\mu_{1,C}$. To identify the energy contributions present in Eq. (47), we rewrite Eq. (47) as follows:

$$m\frac{d^2\Delta_{tr}}{dt^2} = -F_{e,tr} + F_{a,tr} \tag{52}$$

Here, $F_{e,tr}$ represents linearized, parametric electromagnetic force, and $F_{a,tr}$ represents the aeroelastic force. Multiplying Eq. (52) by the velocity, $\dot{\Delta}_{tr}$ gives:

$$\frac{d}{dt}\frac{1}{2}m\dot{\Delta}_{tr}^2 = -F_{e,tr}\dot{\Delta}_{tr} + F_{a,tr}\dot{\Delta}_{tr} = P_e + P_a$$
$$P_e = -F_{e,tr}\dot{\Delta}_{tr} \tag{53}$$
$$P_a = F_{a,tr}\dot{\Delta}_{tr}$$

Here, $P_e$ is the power input by the parametric electromagnetic force, and $P_a$ is the power input by the aeroelastic force. Now, by integrating the energy variation law over one cycle, i.e., from $0$ to $t_{end} = 2\pi/\Omega$, we obtain the energy balance:

$$\int_0^{t_{end}}\frac{d}{dt}\frac{1}{2}m\dot{\Delta}_{tr}^2 dt = \int_0^{t_{end}}P_e dt + \int_0^{t_{end}}P_a dt \Rightarrow E_{kin}(t_{end}) - E_{kin}(0) = W_e + W_a = W_t \tag{54}$$

In Eq. (54), $E_{kin}$ is the kinetic energy; and $W_e$, $W_a$, and $W_t$ are the electromagnetic, aeroelastic, and total work done per cycle, respectively.

Fig. 7 illustrates the energy analysis performed using Eq. (54) for various aeroelastic forcing scenarios (discussed in Fig. 5). In panel (a), point 1 (inside black ellipse) from Fig. 5 is considered, with $\mu_{1,C} = 0$, $\mu_{1,S} = 0$, and $\mu_0$ is then slightly increased to study its effect on the work done by the forces (at point 1). At $\mu_0 = 0$ we have $W_t > 0$, which confirms that steady-state equilibrium is unstable. As $\mu_0$ increases, $W_a$ also increases, clearly indicating that the aeroelastic force with constant coefficient further destabilizes the steady-state equilibrium. The electromagnetic force, however, transitions from being destabilizing to stabilizing as $\mu_0$ increases, as the controller attempts to nullify the destabilizing effects (although not successfully, as $W_t > 0$ for all $\mu_0$ considered). Notably, around $\mu_0 = 0.4$, the work done by the electromagnetic force becomes zero.

In panel (b) of Fig. 7, point 2 (inside red ellipse) in Fig. 5 is considered, with $\mu_0 = 4000$, $\mu_{1,S} = 0$. Now, the influence of increasing $\mu_{1,C}$ on the work done by the forces at point 2 is studied. As $\mu_{1,C}$ increases, $W_a$ decreases, suggesting that the parametric aeroelastic force has a stabilizing character, in contrast to the destabilizing character of the aeroelastic force with constant ($\mu_0$) coefficient. Starting from the situation where the left vertex of the red ellipse lies at point A (see Fig. 5), the ellipse shrinks as $\mu_{1,C}$ increases and the left vertex passes through point 2. Further increase leads to a complete collapse of the ellipse (when $\mu_{1,C}$ reaches $\mu_{opt} = 8000$; see Eq. (49)), and the left vertex (as well as the right one and the centre) end up at point B. Beyond this point, the natures of the forces reverse. In contrast to the initial behaviour, the (total) aeroelastic force now exhibits a destabilizing character and the parametric electromagnetic force a stabilizing one. The ellipse then begins to grow again, ultimately returning to the initial shape with the left vertex at point C (which is the same as point A, but the value of $\mu_{1,C}$ is different).

In panel (c) of Fig. 7, $\mu_0 = 0$, $\mu_{1,S} = 0$ are chosen to investigate the influence of the parametric aeroelastic force separately, and again point 1 (inside black ellipse) is considered; $\mu_{1,C}$ is increased (to study its effect on the work done by the forces at point 1). Like observed in panel (b), the parametric aeroelastic force shows a dual character, as it can again be either stabilizing or destabilizing; however, the dual character can now be unambiguously attributed to the parametric aeroelastic force as the aeroelastic force with constant coefficient is simply absent. Throughout the stable region (and even outside it, to its left), $W_a$ remains negative, indicating that the parametric aeroelastic force is stabilizing, ultimately leading to the disappearance of the ellipse (when $\mu_{1,C}$ reaches $\mu_{opt} = 8000$; see Eq. (49)). In contrast, in panel (b), $W_a$ initially has a positive value even within the stable region, which is now observed to result from the influence of $\mu_0$.

Finally, in panel (d), a simple case is shown, with $\mu_0 = 0, \mu_{1,C} = 0, \mu_{1,S} = 0$ and we observe point 3 (outside black ellipse, to its left) in Fig. 5. Now, the amplitude of the base excitation, $A$ (see Eq. (32)), is varied to study its effect on the work done by the forces at point 3. It is observed that $W_e$ transitions from negative to positive as the left vertex of the black ellipse crosses point 3, which shows how the controller fails to stabilize beyond a particular value of $A$.

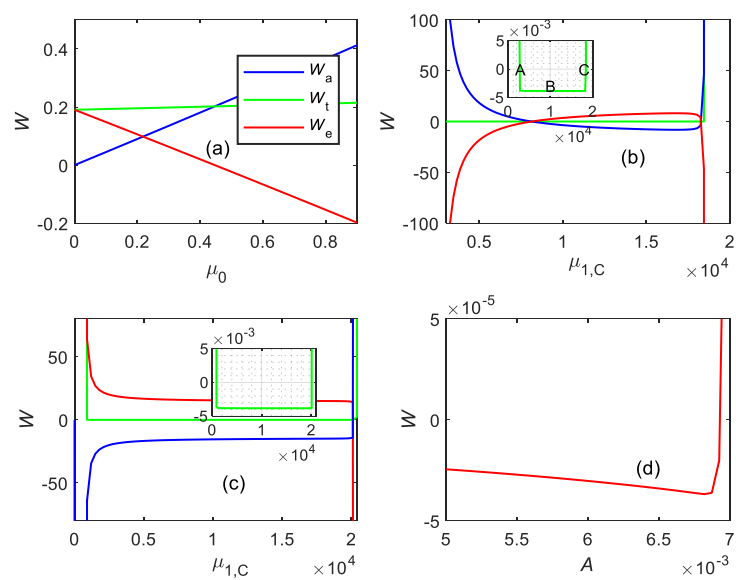

Figure 7. In (a), $\mu_{1,C}=0, \mu_{1,S}=0, K_p=19500(\text{V/m})$ and the destabilizing character of the aeroelastic force with constant coefficient is shown. In (b), $\mu_0=4000, \mu_{1,S}=0, K_p=16000(\text{V/m})$ and the interplay between $\mu_0$ and $\mu_{1,C}$ is shown. In (c), $\mu_0=0, \mu_{1,S}=0, K_p=19500(\text{V/m})$ and the dual character of parametric aeroelastic force, without the influence of aeroelastic force with constant coefficient, is shown. In (d) $\mu_0=0, \mu_{1,C}=0, \mu_{1,S}=0, K_p=19000(\text{V/m})$ and the influence of $A$ on $W_e$, without the influence of aeroelastic force is shown. The remaining parameter values are given by, $K_d=10000(\text{Vs/m})$, $C=0.05(\text{Nm}^2/\text{A}^2)$, $z_0=0.015(\text{m})$, $m=7650(\text{kg})$, $R=9.71(\text{Ohm})$, $A=0.0142(\text{m})$, $\omega=40(\text{rad/s})$.

From Fig. 7 panels (a)-(c), we can draw the following conclusions. Adding the state-dependent aeroelastic force (with a constant coefficient) alters the energy input by the parametric electromagnetic force (Fig. 7 panel (a)), which is state-dependent too. That aeroelastic force even changes the character of the parametric electromagnetic force from destabilizing to stabilizing if $\mu_0$ is sufficiently large, although it does not change of the stability of the equilibrium (Fig. 7 panel (a)). The parametric aeroelastic force can, however, stabilize the equilibrium if $\mu_{1,C}$ is sufficiently large (Fig. 7 panel (b) and panel (c)); note that the parametric aeroelastic force can even completely eliminate the parametric resonance throughout the $K_p - K_d$ plane if $\mu_{1,C}$ is chosen appropriately, as shown before in Fig. 5. The parametric aeroelastic force can (like the aeroelastic force with constant coefficient) change the character of the parametric electromagnetic force from destabilizing to stabilizing, depending on $\mu_{1,C}$; however, the overall stability transition induced by the parametric aeroelastic force (the aeroelastic force with constant coefficient cannot do that) can be verified not to coincide (i.e., lies at another $\mu_{1,C}$ value) with the transition in the character of the parametric electromagnetic force (Fig. 7 panels (b) and (c)).

## 5 Conclusions

This study analyses the stability of a 1.5-degree-of-freedom model consisting of an electromagnetically suspended mass that is excited by an oscillating base and the aeroelastic force. The model is a simplified representation of a Hyperloop vehicle moving through air.

For the case without external excitation (i.e., no oscillating base), analytical expressions were derived for the stability boundaries by employing linear stability analysis. The results indicate that the control parameter space ($K_p - K_d$) is divided into three distinct regions, one of which exhibits limit cycle behaviour, akin to that beyond the supercritical Hopf bifurcation. The presence of the aeroelastic force (with constant coefficient) leads to a marginal reduction of the size of the stable region, with no qualitative changes in the stability landscape. Harmonic balance analysis identified the region in the control parameter space where the limit cycle exists as well as its amplitude and frequency. The present study considers a PD-controller. However, the same approach can be used for more complicated controllers such as PID to identify the stability boundaries.

For the case with base excitation, the stability boundaries were also determined, both analytically and numerically. The right boundary now consists of the same inclined line as obtained for the non-oscillating base scenario as well as an ellipse related to parametric resonance located on it, indenting the stable domain. The inclined line is influenced by a aeroelastic force with constant coefficient, while the size of the elliptical region is affected by the aeroelastic force with harmonically varying coefficient (i.e., the parametric aeroelastic force, which is added on top of the one with constant coefficient). The study reveals the possibility of eliminating parametric resonance induced by one parametric force by adding another, with the phase shift between these two parametric forces being crucial and ideally equal to zero. Energy analysis demonstrates that if the parametric aeroelastic force has a nonzero but small phase difference with the base excitation, the interplay between electromagnetic and aeroelastic parametric forces can still result in the suppression of parametric resonance. Finally, the results of this paper are applicable to similar electromagnetic [18] and other systems for mitigating parametric resonance caused by one state-dependent force using another.

## Acknowledgement

The authors express sincere gratitude to the European Union's Horizon Europe programme for its support through the Marie Skłodowska-Curie grant agreement No 101106482 (HySpeed project).